\documentclass[prb,twocolumn,showpacs,preprintnumbers,amsmath,amssymb]{revtex4}

\usepackage{graphicx}
\usepackage{dcolumn}
\usepackage{bm}

\begin{document}


\title{Nonharmonic oscillations of nanosized cantilevers due to quantum-size effects    }

\author{Martin Olsen}
\email{Martin.Olsen@miun.se}
 \author{Per Gradin}
 \author{Ulf Lindefelt}
\author{H\aa kan Olin }
\email{Hakan.Olin@miun.se}
 \affiliation{Department of Natural Sciences, Engineering and Mathematics, Mid Sweden University,  SE-851 70 Sundsvall, Sweden}
 

\date{\today}

\begin{abstract}
   Using a one dimensional jellium model and standard beam theory we calculate the spring constant of a vibrating nanowire cantilever.
   By using the asymptotic energy eigenvalues of the 
   standing electron waves over the nanometer sized cross section area, the change in the grand canonical potential is calculated and hence the force and
   the spring constant. 
    As the wire is bent more electron states fits in its cross section.
    This has an impact on the spring "constant" which oscillates slightly with the bending of the wire. 
    In this way we obtain an amplitude dependent resonance frequency of the oscillations that should be detectable.
   \end{abstract}

\maketitle

\section{\label{sec:level1} Introduction}
   Nanoelectromechanical (NEMS) systems are usually still in the macroscopic regime in the sense that quantum effects do not play a large role. This is due 
   to the smallness of the quantum energy of a mechanical structure even if it is of nanometer dimensions. For example, $\hbar \omega$ for a cantilever with a large 
   resonance frequency of 1 GHz will only have an energy of 4 $\mu$eV, which is much smaller than the termal energy at normal conditions. The corresponding
   amplitude $A$ of such quantum oscillations, given by  $ \frac{ 1 }{2 } k A^2 = \hbar \omega$, where k is the transverse spring constant, 
   will be very low and hard to detect \cite{Blencowe:2003}.
   However, in certain circumstances quantum size effects have a significant effect \cite{Agrait:3003}. For example, 
    Stafford \emph{et al.} \cite{Staff:1997} and others \cite{Ruiten:1997,Yannoul:1997,Blom:1998},  calculate the tensile force in a
   nanowire during its elongation.  They found jumps in the force corresponding to different numbers of electron states that fits in the wires cross section when the wire was stretched. 
   This effect has been measured \cite{Rubio:1996,Stadler:1996}.
    The canonical component in NEMS is the cantilever, which is used in a large number of systems including atomic force microscopes and cantilever based sensors.
    It is not obvious that bending a cantilever will give the same effect as in the Stafford system where the diameter of the nanowire could be 
    reduced to a fraction.        
    
    Here, we show that quantum size effects should be included for thin cantilevers, and even if the effect is small, at resonance the effect will be detectable
    due the the high accuracy at which frequency of oscillation can be measured: down to tenth of mHz.\cite{Giessibl:2003}
    We use the same kind of free electron model as in earlier studies to calculate the spring constant of the cantilever.
   \section{\label{sec:level1} Model} 
\begin{figure}
   \centering\resizebox{0.5\textwidth}{!}{\includegraphics{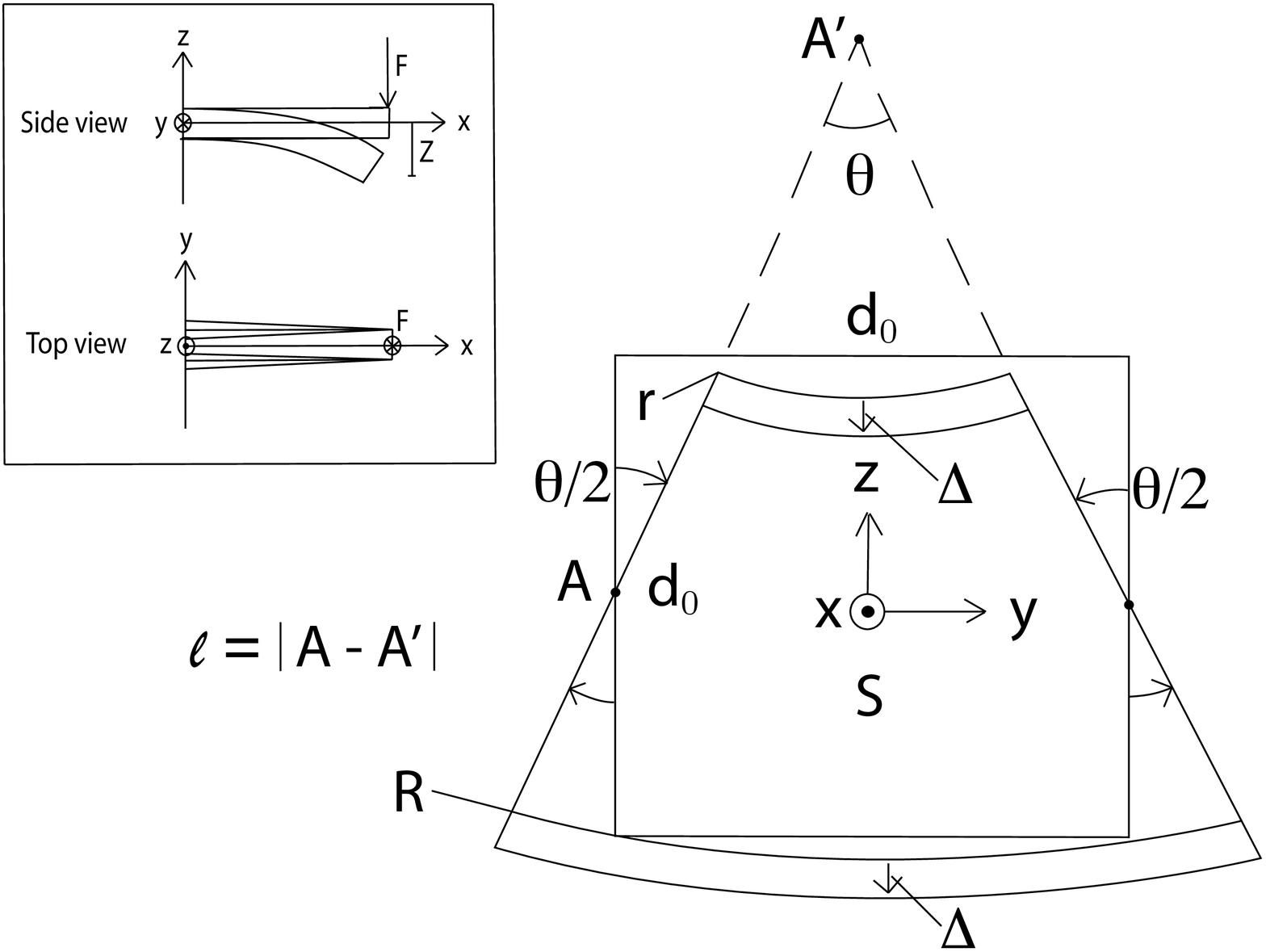}}
   \caption{A cross section $S$ of the cantilever before and after a perpendicular force is applied on the wires end. 
   Inset: Side view and top view of the cantilever nanowire. The expansions and contractions of the cross sections are larger nearer 
   the fixed end if the cross section can contract and expand freely.}
   \label{figdefyta}
   \end{figure}

    When a nanowire is bent the length of a fibre on the upper side of the wire increases, corresponding to reduction of the width of the cross section. Conversely
    the length is reduced on the lower part corresponding to an increase of the width of the cross section. 
    In the middle is an unaffected neutral line.
    This effects will change the cross sectional area of the nanowire which in turn change the tensile force in the wire as it becomes more bent.
    Because the wire is assumed to be thin, only a few wave modes under the Fermi level fits in the cross section.    
   We use a one dimensional density of states in a similar way as in earlier works \cite{Staff:1997,Ruiten:1997,Yannoul:1997,Blom:1998}.
   We consider a straight nanowire, which we bend by applying a perpendicular force at the free end. The other end of the wire is attached
   to bulk material letting electrons flow in and out of the wire. By using standard beam theory we find that bending of the wire yields an increase in the cross
   section area.
    The wire has the length $L$ and a quadratic cross section with undeformed side $d_{0}$, see fig~\ref{figdefyta}.
   When we applying a perpendicular force $F$ at the end of the cantilever at $x = L$, we obtain 
   \begin{eqnarray} 
      M &=& F  \left( L-x \right) ,  \\
      \sigma_{x} &=&  \frac{M }{I} z  ,  \\
      \sigma_{y} &=& \sigma_{z} = 0, 
   \end{eqnarray}
    where $M$ is the moment of the force $F$ at position $x$, $I$ is the areal moment of inertia of the cross section and $\sigma_{x}, \sigma_{y}, \sigma_{z}$
    are stresses in the $x$, $y$ and $z$ direction respectively. For strains in the $x$, $y$ and $z$ directions we have
   \begin{eqnarray}
     \varepsilon_{x} &=& \frac{\sigma_{x} }{E}  , \\
      \varepsilon_{y} &=&  - \nu \varepsilon_{x}  ,  \\
      \varepsilon_{z} &=& - \nu \varepsilon_{x}   , 
   \end{eqnarray}
   where $\nu$ is the Poisson's ratio and $E$ is the Young's modulus.
   This lead to the deformation of the cantilever cross section as shown in fig~\ref{figdefyta}.
   For the deformations $u_y$ we have
   \begin{eqnarray} 
      \varepsilon_{y} &=& \frac{ \partial u_y }{ \partial y }   \Rightarrow   \\
      u_y &=& - \frac{  \nu F }{E I }  \left( L-x \right) y z + H  \left( x, z \right)  ,
   \end{eqnarray}
   using partial integration, where $H  \left( x, z \right) \equiv 0$ by symmetry.  For the deformations $u_z$ we have
   \begin{eqnarray} 
      \varepsilon_{z} &=& \frac{ \partial u_z }{ \partial z }   \Rightarrow   \\
      u_z &=& - \frac{  \nu F }{2 E I }  \left( L-x \right)  z^2 + G  \left( x, y \right)  .
   \end{eqnarray}   
   The shear strain $\gamma_{yz }$ is assumed to be zero, so
    \begin{eqnarray} 
       \gamma_{yz } &=& \frac{ \partial u_y }{ \partial z }  + \frac{ \partial u_z }{ \partial y }  = 0   ,
    \end{eqnarray}
    yielding an expression for $G  \left( x, y \right)$ to use in equation (10). We finally obtain
    \begin{eqnarray} 
      u_y &=& - \frac{  \nu F }{E I }  \left( L-x \right) y z   ,  \\
      u_z &=& - \frac{  \nu F }{2 E I }  \left( L-x \right)   \left( z^2 - y^2  \right)+ f  \left( x \right)  ,
   \end{eqnarray}
   where $f  \left( x \right)$ is the engineering beam theory solution describing the bending of the neutral line $y=z=0$ which only implies a translation of the whole cross section
   and therefore can be left out when calculating its deformation.    
   
   When the wire is deformed by the force, the side of the cross section is bent inwards with an angle $\theta/2$ on the upper part of the wire cross section and is bent 
   outwards with the same angle on the lower side. On the top side there is a compression of the wire and on the down side there is an elongation.
    The area after the deformation is the difference between two circle sectors with radii $R$ and $r$ respectively and $ \theta$ is the top angle:
    \begin{eqnarray} 
      S &=& \pi  \left( R^2  - r^2 \right)  \frac{ \theta  }{2 \pi} ,  \\
      R &=& \ell +  \frac{ d_{0} }{2} + \Delta ,  \\
       r &=&  \ell -  \frac{ d_{0} }{2} + \Delta ,  \\
       \sin \left( \frac{ \theta }{2} \right)  &=& \frac{ d_0  }{ 2 \ell }  ,
   \end{eqnarray}
   resulting in $S \approx  d_0^2  \left( 1+   \Delta /   \ell      \right)$. 
   From (13) we see that
     \begin{eqnarray} 
      \Delta  &=&  \frac{  \nu F }{2 E I }  \left( L-x \right)  \left(   \frac{  d_0 }{2 }   \right)^2  .
   \end{eqnarray}
   The radius of curvature $\ell$ is given by
   \begin{eqnarray} 
       \frac{  1 }{ \ell  } &\approx& \rvert   \frac{ \partial^2 u_z }{ \partial y^2 }  \rvert  =   \frac{  \nu F }{ E I }  \left( L-x \right)   .
   \end{eqnarray}   
   We then obtain using (18) and (19):
   \begin{eqnarray} 
       S = d_{0}^2  \left[1+   \frac{ 1}{8}  \left(   \frac{  \nu  F  \left( L-x \right) d_{0} }{ E I}  \right)^2   \right]   .
   \end{eqnarray} 
   A beam of length $L$ with a perpendicular force $F$ applied at the end bends a distance $Z$:
   \begin{eqnarray} 
        Z &=& \frac{ F L^3 }{3 E I } = \frac{ k Z L^3 }{3 E I }  ,
   \end{eqnarray}
   where $F = k Z$, and $k$ in turn is the transverse spring constant of the wire. We then obtain
   \begin{eqnarray} 
       S = d_{0}^2  \left[1 +   \frac{ 9 \nu^2 d_{0}^2  \left( L-x \right)^2  Z^2  }{8 L^6}      \right]  ,
   \end{eqnarray}
   which is independent of the Young's modulus $E$.
   The eigenenergies for the standing electron waves that fits the cross section is in the limit of large eigenvalues \cite{Strauss:1992}
   \begin{eqnarray} 
       E_{n} =    \frac{ \hbar^2  }{2m}  \frac{ 4 \pi  }{S} n  ,
   \end{eqnarray}
   where $n$ is the quantum number and $m$ the electron mass. 
   The grand canonical potential of the electron gas in the nanowire is for low temperature given by \cite{Staff:1997,Blom:1998}
   \begin{eqnarray}
       \Omega &=& - \sum_{n}   \int_{0}^{L  }  \frac{ 4 }{3 }   \sqrt{ \frac{ 2 m }{\pi^2 \hbar^2   }  }
        \left(  E_{F} - E_{n}\left(   x   \right)   \right)^ {3/2} dx  .
   \end{eqnarray}  
   Integration of (24) using (22) for  \emph{small} bending $Z$ yielded 
   \begin{eqnarray}
       \Omega   =  - \sum_{n}      \sqrt{ \frac{ 2 m }{\pi^2 \hbar^2   }  }    \left[    \frac{ 4 }{3 } \left(  E_{F} - E_{n0}   \right)^ {3/2} L \right.    \phantom{22222222   }    \nonumber  \\
                        \left.      +  \frac{ 2 }{3 }   \sqrt{ E_{F} - E_{n0}  } E_{n0}  \frac{ 9 }{8 }  \left(    \frac{  d_{0} \nu  }{ L^3} \right)^2 Z^2 L^3   \right] ,
   \end{eqnarray} 
   which is Blom \emph{et al.} \cite{Blom:1998} plus a term proportional to the down bending $Z$ squared. We have
   \begin{figure}
   \centering\resizebox{0.5\textwidth}{!}{\includegraphics{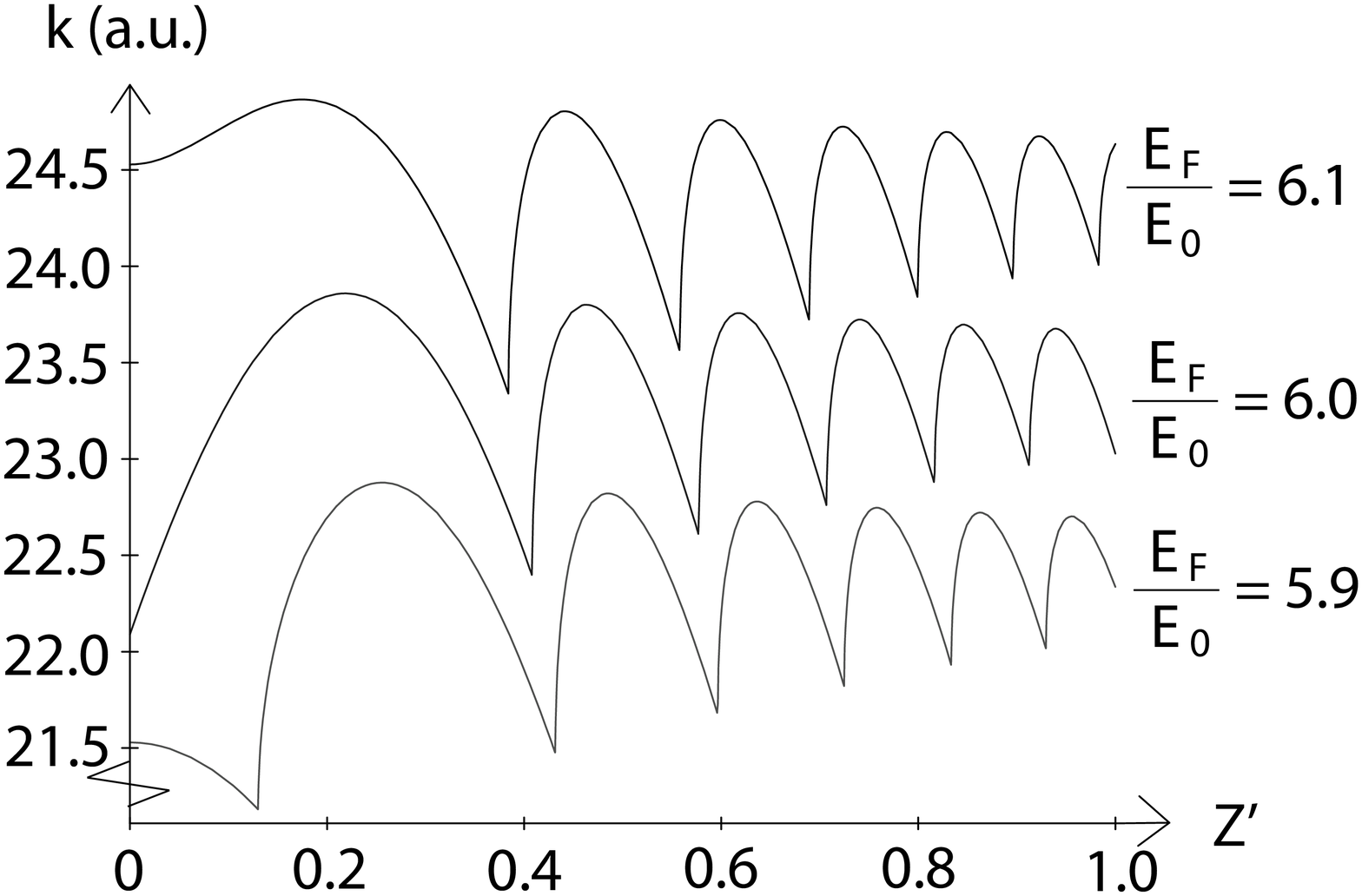}} 
   \caption{Example of spring constant $k$ variation as a function of bending $Z' = \sqrt{   \frac{  9  }{  8   }  }  \frac{  \nu d_{0}    Z  }{ L^2}$
    obtained from an  equation having the same type of dependency  on $n$ and $Z$ as (32) has, i.e. equation (38). 
    The three curves show the forms of the curve obtained from (38) for $\frac{  E_F }{ E_0 }$ below, at and above an integer.
    We see that (38) is independent of the sign of $Z$ as is to be expected by symmetry.
   The k-axis always intersects the curve at a local maxima or minima. However, this maxima or minima may not be very wide depending on
   how close we are that a new electron state will fit when bending the wire slightly. We see in the figure that the width is small when $\frac{  E_F }{ E_0 }$ is close to an integer 
   (middle curve) and broader farther from an integer.}
   \label{fjad}
   \end{figure}  
   \begin{eqnarray} 
       E_{n0} &=&    \frac{ \hbar^2  }{2m}  \frac{ 4 \pi  }{d_{0}^2} n = E_{0} n ,  \\
       N &=&    \frac{ E_F  }{E_0} ,
   \end{eqnarray}
   where $N$, if rounded off down to an integer, is the number of energy levels below Fermi level.
   Using for example $d_{0} = 4$ nanometer and $E_{F} = 5.5$ electron volt (gold, silver) we obtain the number of energy levels below Fermi level $N=183$. $d_0=1$ nm
   yields $N=11$.
   The force $F$ which is assumed to be due to the electron gas is given by $F =  - \frac{ \partial   \Omega }{ \partial Z } = k Z$, so we obtain, assuming
   $ \frac{ \partial L }{ \partial Z } = 0$  (no elongation of the wire), the spring constant
   \begin{eqnarray}
       k_0  &=& \sum_{n=1}^N   \frac{ 3 d_{0} ^2 \nu^2   }{  2 L^3}   \sqrt{ \frac{ 2 m }{\pi^2 \hbar^2   }  }   \sqrt{ E_{F} - E_{n0}  }  E_{n0}   . 
   \end{eqnarray}
   Disregarding the variation of $S$ with $x$ in (22) and assuming the same deformation in every cross section as at the fixed end to simplify the integration we obtain
    \begin{eqnarray} 
       S = d_{0}^2  \left[1 +    \frac{9  \nu^2 d_{0}^2    Z^2  }{8 L^4}      \right]  .
   \end{eqnarray}
   Using (29) in (24) and taking the derivative yielded
    \begin{eqnarray}
       k  = \sum_{n=1}^N        \sqrt{ E_{F} -    \frac{ E_{0}  n    }{  1+ \frac{ 9 \nu^2 d_{0}^2    Z^2  }{  8 L^4}   }   }  
       \frac{   \sqrt{   \frac{ 2 m }{\pi^2 \hbar^2   }  }    n E_{0}  \frac{    18 d_{0} ^2 \nu^2  }{ 4 L^3}     }{  \left[   1+ \frac{ 9 \nu^2 d_{0}^2    Z^2  }{8 L^4}    \right]^2  }   
   \end{eqnarray}
   where 
   \begin{eqnarray} 
               N &=&    \frac{ E_F  }{E_0} \left(   1+  \frac{ 9 \nu^2 d_{0}^2    Z^2  }{8 L^4}    \right)  .
   \end{eqnarray}
   Because (30) should yield the same as (28) for small bending, (30) should be corrected with factor $1/3$ to account for the variation of $S$ with $x$: 
   \begin{eqnarray}
       k  = \sum_{n=1}^N      \sqrt{ E_{F} -    \frac{ E_{0}  n   }{  1+  \frac{ 9 \nu^2 d_{0}^2    Z^2  }{8 L^4}   }   }  
       \frac{ \sqrt{   \frac{ 2 m }{\pi^2 \hbar^2   }  }   n E_{0}  \frac{    6 d_{0} ^2 \nu^2  }{ 4 L^3}     }{  \left[   1+  \frac{ 9 \nu^2 d_{0}^2    Z^2  }{8 L^4}    \right]^2  }   
   \end{eqnarray}
   This expression is valid for \emph{arbitrary} bending $Z$.
   Plotting $k$ from equation (32) as a function of $Z$ is shown in fig~\ref{fjad}.
   Replacing a sum by an integral we found 
   \begin{eqnarray} 
       \sum_{n=1}^{N} n \sqrt{ N - n  }  &\approx&  \frac{ 4 }{15   }  \left(  N^{5/2} - N     \right)  .
   \end{eqnarray}  
    Using (28) for  \emph{small} bending and (33), neglecting the second term in the RHS of (33) because $N > 1$, we obtain the spring constant
    for the unbent wire
   \begin{eqnarray} 
      k_0 &=&    \frac{   \sqrt{ 2}   \nu^2 }{5 \pi^2  }      \sqrt{  \frac{  d_{0}^2 m^3 E_{F}^5 }{  \hbar^6  } }   \left( \frac{ d_{0} }{L   }\right)^3   ,
   \end{eqnarray}
   due to an increase in the electron gas density of states that fits in the nanowire when bent. 
      The k-axis always intersect the curve at a local maxima or minima of the k-values, as we see in fig~\ref{fjad}.
    We may therefore have $k=k_0 + const \times Z^2 $ for not too large bending, see (39). 
    \section{\label{sec:level1} Discussion}
    We may then rewrite the harmonic equation with a bending dependent spring constant:
     \begin{eqnarray} 
        \frac{  d^2 Z   }{  d t^2   } + \omega_0^2 Z  + \beta Z^3  + 2 \gamma  \frac{  d  Z   }{  d t   }   = F  \sin \left(   \omega t  \right)  .
   \end{eqnarray}
   This is called the Duffing equation.
   The amplitude $A$ and phase $ \phi$ of the stationary solution to the linear ($\beta =0$) differential equation is given by
    \begin{eqnarray} 
        A &=&   \frac{  F  }{  \sqrt{  \left(  \omega_0^2  -  \omega^2   \right) ^2       +  \left(  2    \omega  \gamma   \right) ^2 }   }  ,  \\
        \tan   \phi &=& - \frac{  2    \omega  \gamma   }{    \omega_0^2  -  \omega^2   }   .
   \end{eqnarray} 
   \begin{figure}
   \centering\resizebox{0.5\textwidth}{!}{\includegraphics{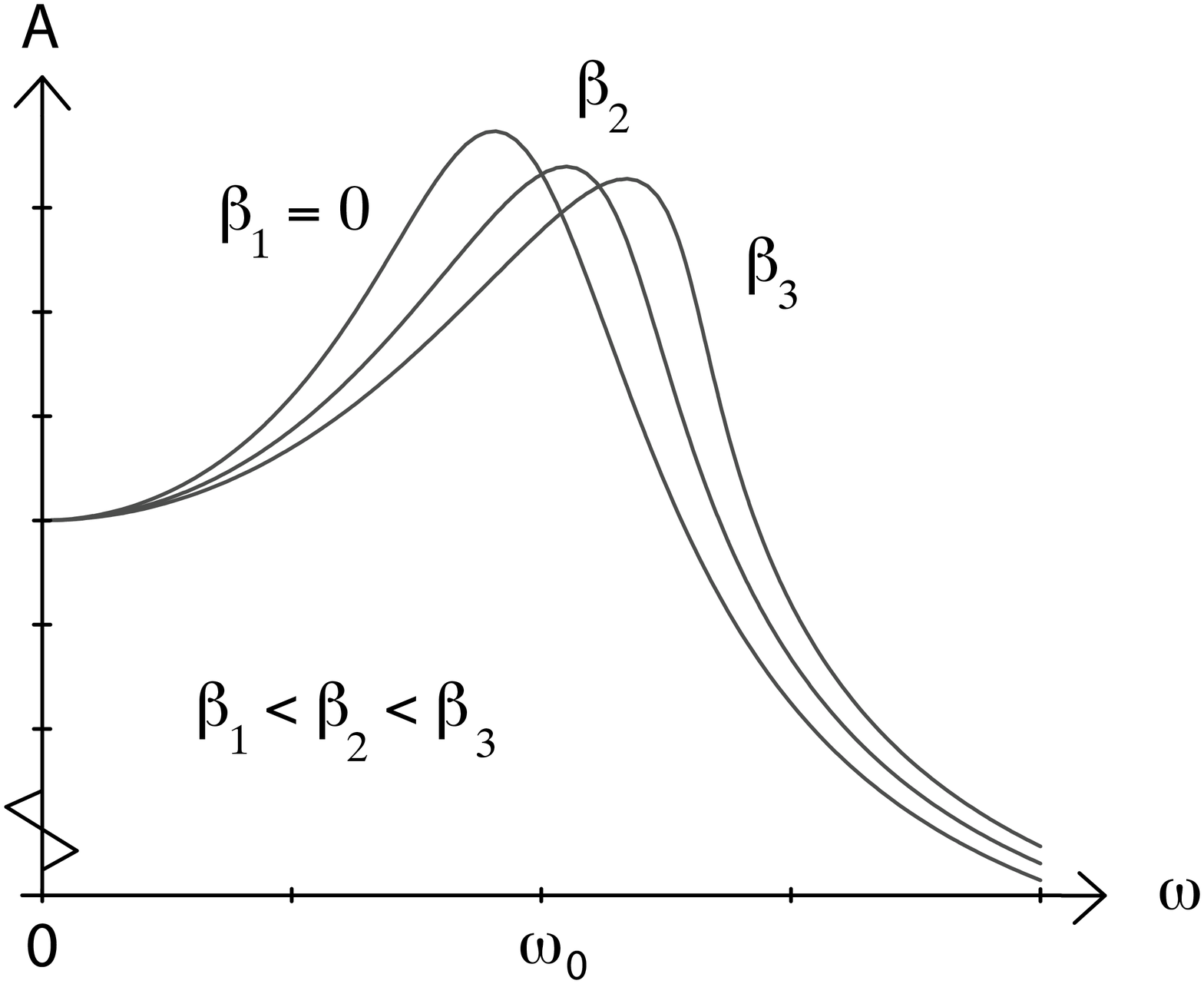}} 
   \caption{The amplitude response of a driven weakly nonlinear harmonic oscillator.  $\beta = 0$ corresponds to the harmonic case where
   the resonance frequency is $ \omega_0$. 
     When the amplitude is increased (or $\beta$ is increased), the resonance frequency is shifted towards higher values for positive $\beta$
     and towards lower values for negative $\beta$, from $ \omega_0$ to $  \sqrt{ \omega_0^2 +   \frac{  3  }{  4   }  \beta A^2 } $. }
   \label{Ampl}
   \end{figure} 
   Because the weak non-linearity the resonance frequency is shifted from $ \omega_0$ to $  \sqrt{ \omega_0^2 +   \frac{  3  }{  4   }  \beta A^2 } $  ,
    see reference \cite{Nayfeh:1979}. 
   We then obtain a shift in the amplitude maximum towards a higher frequency for positive $\beta$, as shown in fig~\ref{Ampl}. 
   Because frequency can be measured at high precision, even small changes can be experimentally detected.  
   The oscillation in the weakly non-linear case takes place around the same point of equilibrium as in the low amplitude, i.e. 
   linear harmonic case. 
   
   We see from fig~\ref{fjad} that $\beta$ is positive for the upper curve and negative for the lower curve around $Z=0$. 
   The middle curve is problematic: very close to $E_F/E_0$ being an integer we have $k=k_0+const  \times  \rvert Z  \rvert$. This discontinuity in the
    first derivative of the curve at $Z=0$ vanishes however quickly when we move away from $E_F/E_0 = integer$.
   
    The results in this paper should be valid for wires with small enough diameter so that the one dimensional distribution function
     is a good approximation.
    From the result presented, it should be possible to experimentally determine how close we are to that a new state would fit in the nanowire.
    By first measuring the low amplitude frequency of the oscillating nanowire and then increase the amplitude it should be possible to determine if the
    resonance frequency is increased or decreased corresponding to positive or negative $\beta$.
  \begin{figure}
   \centering\resizebox{0.5\textwidth}{!}{\includegraphics{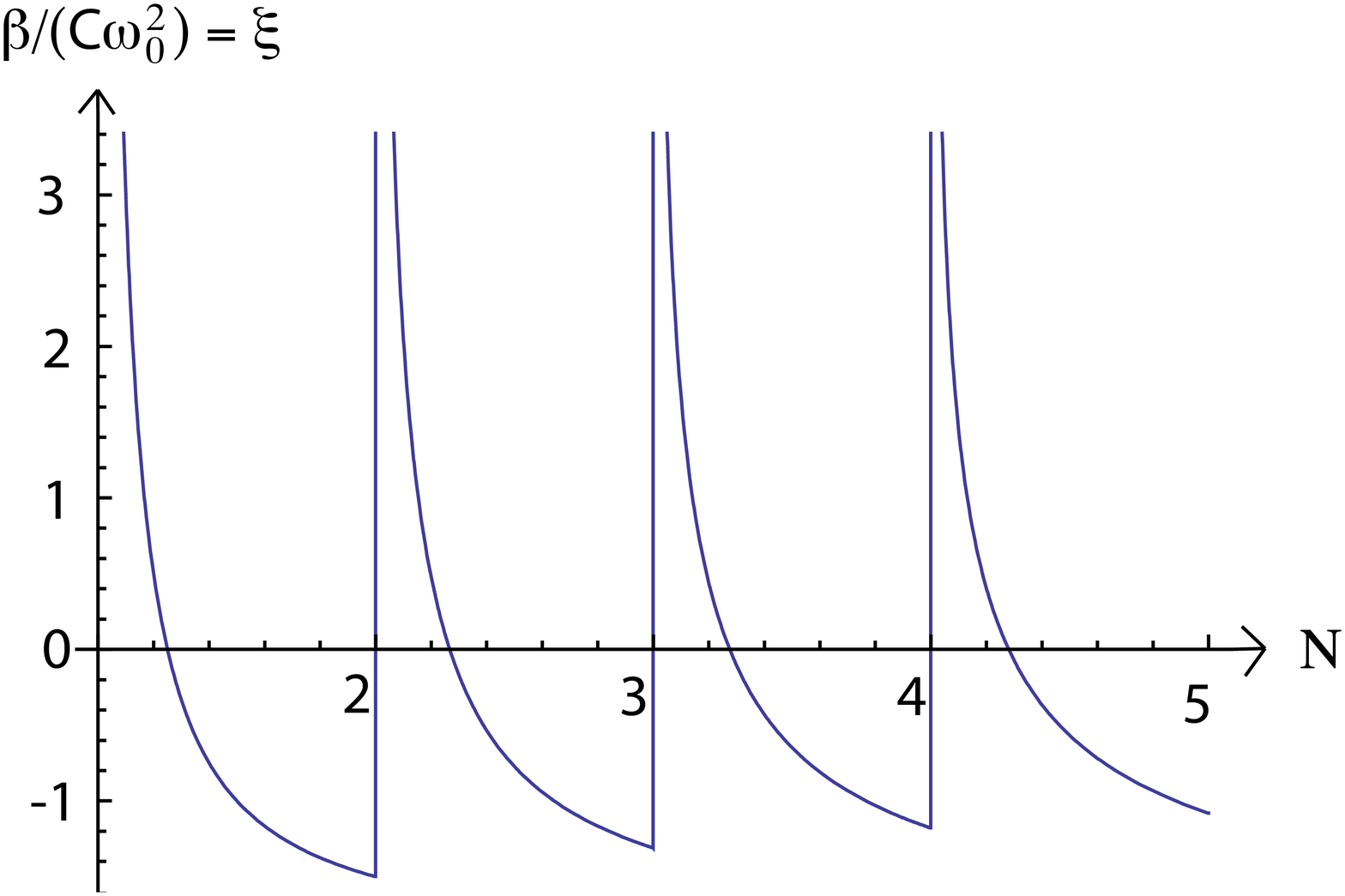}} 
   \caption{ $\beta  / ( C \omega_0^2)$ as a function of $N = \frac{ E_F  }{ E_0  }$ obtained by dividing
   the second term in (39) with the first term and identifying the spring constant in equation (35). $C= \frac{ 9 \nu^2 d_{0}^2  }{8 L^4}$.
   We note the singularities at integer numbers, corresponding to when we have 
   $k=k_0+const  \times  \rvert Z  \rvert$ instead of  $k=k_0 + const \times Z^2 $ which is the normal behaviour near $Z=0$. }
   \label{betaome}
   \end{figure}  
   To illustrate the changes in equation (32) with different values of $N= E_F/E_0$ we made plots. The function to be plotted is
    \begin{eqnarray} 
      k' \left(Z \right)  =  \sum_{n=1}^{N\left( 1+C Z^2 \right)}   \sqrt {  N -      \frac{  n  }{  1+C Z^2   }  }  \times  \frac{  n  }{ \left( 1+C Z^2 \right)^2  }   ,
   \end{eqnarray}
    where $C= \frac{ 9 \nu^2 d_{0}^2  }{8 L^4}$, which has the same type of dependency on $n$ and $Z$ as equation (32) has.
    We then obtain curves as in fig~\ref{fjad}. The lowest curve is obtained for $N=5.9$, the middle curve for $N=6.0$  and the upper curve for $N=6.1$.
   For small bending $Z$, (38) becomes
    \begin{eqnarray} 
     k' \left(Z \right)   &=&  \sum_{n=1}^{N }   n  \sqrt {  N -  n  }   \phantom{  1 } +  \nonumber        \\
      &+& C Z^2   \sum_{n=1}^{N }  \left(    \frac{ 5 n^2 - 4 N n  }{ 2  \sqrt {  N -  n  }   }   \right)  + \ldots  \phantom{  1 } .
   \end{eqnarray}
   Dividing (38) with the first term in (39) yields the relative size of the effect of bending on the spring constant.
   A typical value of the relative change in spring constant due to bending was about 
    $0.01- 0.1 \%$ up to $N=200$ at $ \sqrt{   \frac{  9  }{  8   }  }  \frac{  \nu d_{0}    Z  }{ L^2}  = 0.05$.
   At high $N$ we would need a smaller deflection $Z$ to reach  the local maximum and minimum points in the curve in fig~\ref{fjad}, however the relative change tends to 
   be smaller as $N$ is increased. 
   From (39) and (35) we can calculate $\beta  / ( C \omega_0^2)$ in equation (35). This yields
   \begin{eqnarray}
        \xi &=&  \frac{  \beta  }{  C \omega_0^2 } =  \frac{  \sum_{n=1}^{N }  \left(    \frac{ 5 n^2 - 4 N n  }{ 2  \sqrt {  N -  n  }   }   \right)  }{   \sum_{n=1}^{N }   n  \sqrt {  N -  n  }   }  ,
     \end{eqnarray} 
    A plot of equation (40) is shown in fig~\ref{betaome}. 
    
   What effect has a finite temperature on this result?  
    Following  Blom \emph{et al.} \cite{Blom:1998} we have the grand canonical potential $\Omega = E_{tot} - \mu N_{tot}$ where
     the chemical potential $\mu \approx E_F$ at room temperature and
     \begin{eqnarray} 
        N_{tot} &=&   \sum_n   \int_{E_n}^{ \infty  }  g(E - E_n) f(E )   \phantom{1   } dE  , \\
        E_{tot} &=&   \sum_n   \int_{E_n}^{ \infty  }  g(E - E_n) f(E ) E \phantom{1   }d E  , 
     \end{eqnarray}
      \begin{figure}
   \centering\resizebox{0.5\textwidth}{!}{\includegraphics{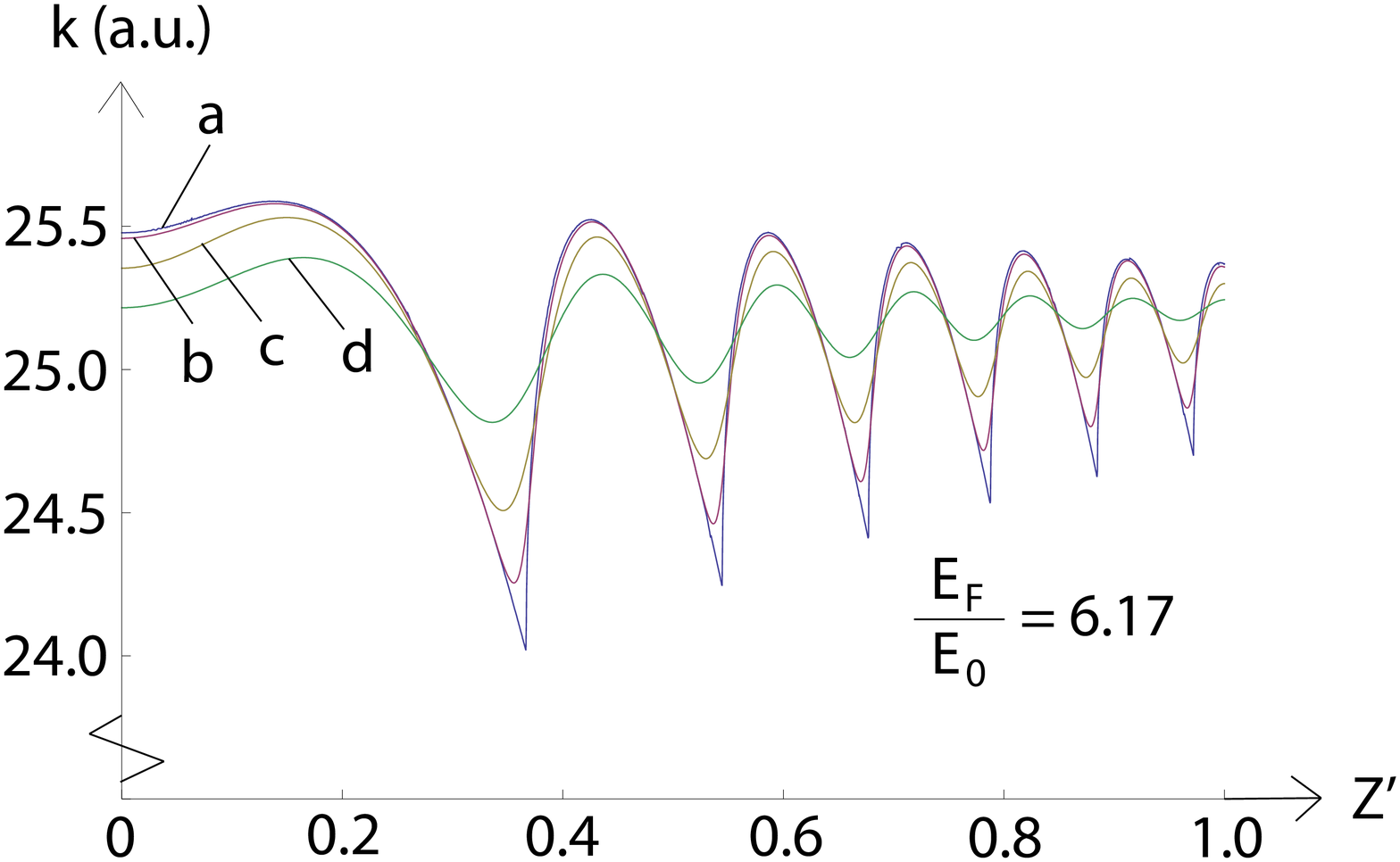}} 
   \caption{Example of spring constant as a function of different bending $Z' = \sqrt{   \frac{  9  }{  8   }  }  \frac{  \nu d_{0}    Z  }{ L^2}$ 
   for four different temperatures using ekvation (43). 
   Curve $a$: $k_B T$ = 0.0001 eV, curve $b$: $k_B T$ = 0.01 eV, curve $c$: $k_B T$ =  0.025 eV (room temperature) and curve $d$: $k_B T$ =  0.05 eV. The plot is made
   using $\mu = 2.9$ eV, $E_0 = 0.47$ eV (i.e. $d_0$ = 1 nm) and the upper energy limit of integration is taken to be $E_{cut} =$ 5.0 eV.
    Increasing $E_{cut}$ to 7.0 yields no visible change in the curves. For $E_{cut} > \mu$ 
   the contribution to the expression decreases rapidly due to the Fermi-Dirac function.   }
   \label{temp}
   \end{figure}  
    where $g(E)$ is the one dimensional density of states, $f(E)$ is the Fermi-Dirac distribution function and $E_n$ is the
    energy of the state $n$ that fits the cross section. Using $F =  - \frac{ \partial   \Omega }{ \partial Z } = k Z$
    and equation (32) we argue that the generalised expression for the spring constant $k$ valid for any temperature should be given by
     \begin{eqnarray}
       k  =   \sum_{n=1}^{  N'  }     \left \{    \frac{ \sqrt{   \frac{ 2 m }{\pi^2 \hbar^2   }  }  
                       n E_{0}  \frac{    6 d_{0} ^2 \nu^2  }{ 4 L^3}     }{  \left[   1+  \frac{ 9 \nu^2 d_{0}^2    Z^2  }{8 L^4}    \right]^2  }  \times   \phantom{22222222222222222   }     \right.  \nonumber    \\
                    \left.   \times    \int_{E_n'}^{ E_{cut}  }     \frac{ 1/2 }{   \sqrt{ E -    \frac{ E_{0}  n   }{  1+  \frac{ 9 \nu^2 d_{0}^2    Z^2  }{8 L^4}   }   }     }   
                        \frac{  d E  }{  \left(   e^{   \frac{  (E - \mu) }{  k_B T   }   }  + 1  \right) }    \right   \}     ,
    \end{eqnarray}
    where 
    \begin{eqnarray}
         N' &=&    \frac{ E_{cut} }{E_0} \left(   1+  \frac{ 9 \nu^2 d_{0}^2    Z^2  }{8 L^4}    \right)  ,  \\
          E_n' &=& \frac{ E_{0}  n   }{ \left( 1+  \frac{ 9 \nu^2 d_{0}^2    Z^2  }{8 L^4}   \right)  }  ,
   \end{eqnarray}
   and the upper integration limit $ E_{cut}  \to   \infty$. Equation (43) reduces to (32) when $T \to 0$. A plot of equation (43) for different temperature is shown in fig~\ref{temp}. 
   
   Due to temperature, more states becomes available for the electrons from $E_F$ up to
   about $E_F + k_B T$ and decreasing the number of available states between $E_F$ and about $E_F - k_B T$. This effect tend to make
   the transition at integer numbers in fig~\ref{betaome} (where a new energy level is added) less sharp, smoothing the spikes in $\xi  = \beta  / ( C \omega_0^2)$. 
   We also see in fig~\ref{temp} that at room temperature the curve follows the zero temperature curve well, except around the minima 
   where a new state is added due to bending. The room temperature  curves here becomes smoothed.
   The condition that temperature are not important \emph{other than close to $N$ being an integer} must be that $k_B T  \ll E_0$ where $E_0$ is the 
   difference in energy between two energy levels. Using equation (26) we obtain this condition as
     \begin{eqnarray} 
         \frac{  k_B T }{E_0 }  &=&  \frac{  k_B T m d_0^2 }{  2  \pi  \hbar^2 }   \ll 1  .
     \end{eqnarray} 
     To obtain agreement between the curve for finite temperature and the zero temperature curve around $Z = 0$ we need equation (46) to be fullfilled. 
     Room temperature corresponds to $ \frac{  k_B T }{E_0} = $ 5.3 \% in fig~\ref{temp}. This means that even for a diameter of 1 nm the system needs to be cooled \cite{Roukes:2002}
     if one is to use the zero temperature result. However, interesting results can also be obtained at room temperature, as we see in the curves in fig~\ref{temp}.
           
    Experimentally, a way to detect these amplitude dependent resonance frequencies might be in situ TEM probing where
    one can see the wire while manipulating it. \cite{poncharal:1999,Svensson:2003lr}
    For the thermal vibration of a nanowire we have $ \frac{  1 }{2 } k_0 A^2  = k_B T$. For weak nanowires this amplitude is relatively large at
    room temperature and can be observed.\cite{Treacy:1996}  If one choose to drive the oscillation with an external electric field this
    effect must be taken into account. Small metallic nanosized cantilever has been manufactured.\cite{Nilsson:2004,Lee:2006,Luber:2008}
    
    \section{\label{sec:level1} Conclusions} 
    Using a one dimensional jellium model and standard beam theory we calculate the spring constant of a vibrating nanowire cantilever.
   By using the asymptotic energy eigenvalues of the 
   standing electron waves over the nanometer sized cross section area, the change in the grand canonical potential is calculated and hence the force and
   the spring constant. 
    As the wire is bent more electron states fits in its cross section.
    This has an impact on the spring "constant" which oscillates slightly with the bending of the wire. 
    In this way we obtain an amplitude dependent resonance frequency of the oscillations that should be detectable.
    Because the weak non-linearity the resonance frequency is shifted from $ \omega_0$ to $  \sqrt{ \omega_0^2 +   \frac{  3  }{  4   }  \beta A^2 } $.
    Using (40) we can replace $\beta$ with $\xi C \omega_0^2$.
    We then obtain from this the relative frequency shift
    \begin{eqnarray} 
          \frac{  \Delta \omega }{ \omega_0 }   &\approx&   \frac{  27 }{64 }    \xi \nu^2     \left(  \frac{  d_0 }{L }   \right)^2       \left(  \frac{  A }{L }   \right)^2  , 
     \end{eqnarray} 
     where $A$ is the amplitude of the oscillation and $\omega = 2 \pi f$. 
     The data of some wires possible to use in an experiment is shown in table~\ref{tabellen}.
       
   \begin{table*}
   \caption{\label{tabellen}Table of different $L =$ 40 mn long gold nanowires ($\nu =$ 0.44) that may be used to measure the predicted effects. 
   The maximum temperature $T$ of the wire
   and its diameter $d_0$ are coupled by equation (46), if one wants to use the zero temperature result. However, as seen in fig~\ref{temp} the modification
   of the curves due to temperature are rather small at room temperature and interesting measurements on the system can also be made at this higher temperature. 
    $\xi$ varies periodically with increasing $N=  \frac{  E_F }{E_0 }$ and is obtained from equation (40) for the zero temperature case. 
   Small changes in $d_0$ (yielding $E_0$) can result in large changes in $\xi$ if $N$ is close to an integer. At about 25-30$\%$ of the distance between 
   $N$ being integers $\xi$ becomes zero as it change sign from positive to negative.
   The frequency shift is proportional to the square of the amplitude $A$. In the table we use $A=$ 12 nm, that is 30$\%$ of the wires length $L$. 
   For the thermal vibration of a nanowire we have $ \frac{  1 }{2 } k_0 A^2  = k_B T$. Using this equation for the weakest  nanowire in the table this amplitude 
    becomes 13 nm at room temperature.   }
   \begin{ruledtabular}
   \begin{tabular}{ccccccc}
   &\multicolumn{2}{c}{ $\phantom{22222222222   }$}&\multicolumn{4}{r}{ $\phantom{22222222   }$  }\\
    $d_0$ (nm)&T (K)&$M (\times 10^{-22 }$ kg) &$k_0 ( mN/m)$&$\xi$&$f_0$ (MHz) &$\Delta f$ (kHz)\\ \hline
    1.0&55\footnote{From equation (46) using $\frac{  k_B T }{E_0 } = 1\%$ and $d_0$ in the table. }
    &1.93\footnote{The mass of the wire $M = \rho L d_0^2$ where $\rho$ is the density. }
    &0.0469\footnote{Calculated using equation (34) and the data in the table. }&-0.344\footnote{Calculated using equation (40) and the data in the table. } 
    &$39.2\footnote{From $f_0 =   \frac{  1 }{2 \pi }  \sqrt {\frac{  k_0 }{M } } $ using $k_0$ and $M$ in the table.}$
    &-0.992\footnote{From equation (47) using the data in the table. $A=$ 12 nm. } \\
    1.5&25\footnotemark[1]&4.34\footnotemark[2]&0.238\footnotemark[3]&-0.464\footnotemark[4]&$58.9\footnotemark[5]$&-4.52\footnotemark[6]\\
    2.0&14\footnotemark[1]&7.72\footnotemark[2]&0.750\footnotemark[3]&-0.395\footnotemark[4]&$78.4\footnotemark[5]$&-9.12\footnotemark[6]\\
    2.5&8.8\footnotemark[1]&12.1\footnotemark[2]&1.83\footnotemark[3]&-0.281\footnotemark[4]&$97.8\footnotemark[5]$&-12.6\footnotemark[6]\\
    3.0&6.2\footnotemark[1]&17.4\footnotemark[2]&3.80\footnotemark[3]&-0.107\footnotemark[4]&$118\footnotemark[5]$&-8.35\footnotemark[6]\\
    3.5&4.5\footnotemark[1]&23.6\footnotemark[2]&7.05\footnotemark[3]&-0.200\footnotemark[4]&$138\footnotemark[5]$&-24.7\footnotemark[6]\\
    4.0&3.4\footnotemark[1] &30.1\footnotemark[2]&12.0\footnotemark[3]&-0.198\footnotemark[4]&$159\footnotemark[5]$&-37.0\footnotemark[6]\\
   \end{tabular}
   \end{ruledtabular}
   \end{table*}


\begin{thebibliography}{18}
\expandafter\ifx\csname natexlab\endcsname\relax\def\natexlab#1{#1}\fi
\expandafter\ifx\csname bibnamefont\endcsname\relax
  \def\bibnamefont#1{#1}\fi
\expandafter\ifx\csname bibfnamefont\endcsname\relax
  \def\bibfnamefont#1{#1}\fi
\expandafter\ifx\csname citenamefont\endcsname\relax
  \def\citenamefont#1{#1}\fi
\expandafter\ifx\csname url\endcsname\relax
  \def\url#1{\texttt{#1}}\fi
\expandafter\ifx\csname urlprefix\endcsname\relax\def\urlprefix{URL }\fi
\providecommand{\bibinfo}[2]{#2}
\providecommand{\eprint}[2][]{\url{#2}}

\bibitem[{\citenamefont{Blencowe}(2004)}]{Blencowe:2003}
\bibinfo{author}{\bibfnamefont{M.}~\bibnamefont{Blencowe}},
  \bibinfo{journal}{Phys. Rep.} \textbf{\bibinfo{volume}{395}},
  \bibinfo{pages}{159} (\bibinfo{year}{2004}).

\bibitem[{\citenamefont{Agrait et~al.}(2003)\citenamefont{Agrait, Yeyati, and
  van Ruitenbeek}}]{Agrait:3003}
\bibinfo{author}{\bibfnamefont{N.}~\bibnamefont{Agrait}},
  \bibinfo{author}{\bibfnamefont{A.~L.} \bibnamefont{Yeyati}},
  \bibnamefont{and} \bibinfo{author}{\bibfnamefont{J.~M.} \bibnamefont{van
  Ruitenbeek}}, \bibinfo{journal}{Phys. Rep.} \textbf{\bibinfo{volume}{337}},
  \bibinfo{pages}{81} (\bibinfo{year}{2003}).

\bibitem[{\citenamefont{Stafford et~al.}(1997)\citenamefont{Stafford,
  Baeriswyl, and B{\"u}rki}}]{Staff:1997}
\bibinfo{author}{\bibfnamefont{C.~A.} \bibnamefont{Stafford}},
  \bibinfo{author}{\bibfnamefont{D.}~\bibnamefont{Baeriswyl}},
  \bibnamefont{and}
  \bibinfo{author}{\bibfnamefont{J.}~\bibnamefont{B{\"u}rki}},
  \bibinfo{journal}{Phys. Rev. Lett.} \textbf{\bibinfo{volume}{79}},
  \bibinfo{pages}{2863} (\bibinfo{year}{1997}).

\bibitem[{\citenamefont{van Ruitenbeek et~al.}(1997)\citenamefont{van
  Ruitenbeek, Devoret, Esteve, and Urbina}}]{Ruiten:1997}
\bibinfo{author}{\bibfnamefont{J.~M.} \bibnamefont{van Ruitenbeek}},
  \bibinfo{author}{\bibfnamefont{M.~H.} \bibnamefont{Devoret}},
  \bibinfo{author}{\bibfnamefont{D.}~\bibnamefont{Esteve}}, \bibnamefont{and}
  \bibinfo{author}{\bibfnamefont{C.}~\bibnamefont{Urbina}},
  \bibinfo{journal}{Phys. Rev. B} \textbf{\bibinfo{volume}{56}},
  \bibinfo{pages}{12566} (\bibinfo{year}{1997}).

\bibitem[{\citenamefont{Yannouleas and Landman}(1997)}]{Yannoul:1997}
\bibinfo{author}{\bibfnamefont{C.}~\bibnamefont{Yannouleas}} \bibnamefont{and}
  \bibinfo{author}{\bibfnamefont{U.}~\bibnamefont{Landman}},
  \bibinfo{journal}{J. Phys. Chem. B} \textbf{\bibinfo{volume}{101}},
  \bibinfo{pages}{5780} (\bibinfo{year}{1997}).

\bibitem[{\citenamefont{Blom et~al.}(1998)\citenamefont{Blom, Olin,
  Costa-Kr{\"a}mer, Garcia, Jonson, Serena, and Shekhter}}]{Blom:1998}
\bibinfo{author}{\bibfnamefont{S.}~\bibnamefont{Blom}},
  \bibinfo{author}{\bibfnamefont{H.}~\bibnamefont{Olin}},
  \bibinfo{author}{\bibfnamefont{J.~L.} \bibnamefont{Costa-Kr{\"a}mer}},
  \bibinfo{author}{\bibfnamefont{N.}~\bibnamefont{Garcia}},
  \bibinfo{author}{\bibfnamefont{M.}~\bibnamefont{Jonson}},
  \bibinfo{author}{\bibfnamefont{P.~A.} \bibnamefont{Serena}},
  \bibnamefont{and} \bibinfo{author}{\bibfnamefont{R.~I.}
  \bibnamefont{Shekhter}}, \bibinfo{journal}{Phys. Rev. B}
  \textbf{\bibinfo{volume}{57}}, \bibinfo{pages}{8830} (\bibinfo{year}{1998}).

\bibitem[{\citenamefont{Rubio et~al.}(1996)\citenamefont{Rubio, Agrait, and
  Vieira}}]{Rubio:1996}
\bibinfo{author}{\bibfnamefont{G.}~\bibnamefont{Rubio}},
  \bibinfo{author}{\bibfnamefont{N.}~\bibnamefont{Agrait}}, \bibnamefont{and}
  \bibinfo{author}{\bibfnamefont{S.}~\bibnamefont{Vieira}},
  \bibinfo{journal}{Phys. Rev. Lett.} \textbf{\bibinfo{volume}{76}},
  \bibinfo{pages}{2302} (\bibinfo{year}{1996}).

\bibitem[{\citenamefont{Stadler and D{\"u}rig}(1996)}]{Stadler:1996}
\bibinfo{author}{\bibfnamefont{A.}~\bibnamefont{Stadler}} \bibnamefont{and}
  \bibinfo{author}{\bibfnamefont{U.}~\bibnamefont{D{\"u}rig}},
  \bibinfo{journal}{Appl. Phys. Lett.} \textbf{\bibinfo{volume}{68}},
  \bibinfo{pages}{637} (\bibinfo{year}{1996}).

\bibitem[{\citenamefont{Giessibl}(2003)}]{Giessibl:2003}
\bibinfo{author}{\bibfnamefont{F.~J.} \bibnamefont{Giessibl}},
  \bibinfo{journal}{Rev. Mod. Phys.} \textbf{\bibinfo{volume}{75}},
  \bibinfo{pages}{949} (\bibinfo{year}{2003}).

\bibitem[{\citenamefont{Strauss}(1992)}]{Strauss:1992}
\bibinfo{author}{\bibfnamefont{W.~A.} \bibnamefont{Strauss}},
  \emph{\bibinfo{title}{Partial differential equations: an introduction}}
  (\bibinfo{publisher}{John Wiley and Sons}, \bibinfo{year}{1992}).

\bibitem[{\citenamefont{Nayfeh and Mook}(1979)}]{Nayfeh:1979}
\bibinfo{author}{\bibfnamefont{A.~H.} \bibnamefont{Nayfeh}} \bibnamefont{and}
  \bibinfo{author}{\bibfnamefont{D.~T.} \bibnamefont{Mook}},
  \emph{\bibinfo{title}{Nonlinear oscillations}}, Pure and applied mathematics
  (\bibinfo{publisher}{Wiley}, \bibinfo{year}{1979}).

\bibitem[{\citenamefont{Mohanty et~al.}(2002)\citenamefont{Mohanty, Harrington,
  Ekinci, Yang, Murphy, and Roukes}}]{Roukes:2002}
\bibinfo{author}{\bibfnamefont{P.}~\bibnamefont{Mohanty}},
  \bibinfo{author}{\bibfnamefont{D.~A.} \bibnamefont{Harrington}},
  \bibinfo{author}{\bibfnamefont{K.~L.} \bibnamefont{Ekinci}},
  \bibinfo{author}{\bibfnamefont{Y.~T.} \bibnamefont{Yang}},
  \bibinfo{author}{\bibfnamefont{M.~J.} \bibnamefont{Murphy}},
  \bibnamefont{and} \bibinfo{author}{\bibfnamefont{M.~L.}
  \bibnamefont{Roukes}}, \bibinfo{journal}{Phys. Rev. B}
  \textbf{\bibinfo{volume}{66}}, \bibinfo{pages}{085416}
  (\bibinfo{year}{2002}).

\bibitem[{\citenamefont{Poncharal et~al.}(1999)\citenamefont{Poncharal, Wang,
  Ugarte, and de~Heer}}]{poncharal:1999}
\bibinfo{author}{\bibfnamefont{P.}~\bibnamefont{Poncharal}},
  \bibinfo{author}{\bibfnamefont{Z.~L.} \bibnamefont{Wang}},
  \bibinfo{author}{\bibfnamefont{D.}~\bibnamefont{Ugarte}}, \bibnamefont{and}
  \bibinfo{author}{\bibfnamefont{W.~A.} \bibnamefont{de~Heer}},
  \bibinfo{journal}{Science} \textbf{\bibinfo{volume}{283}},
  \bibinfo{pages}{1513} (\bibinfo{year}{1999}).

\bibitem[{\citenamefont{Svensson et~al.}(2003)\citenamefont{Svensson, Jompol,
  Olin, and Olsson}}]{Svensson:2003lr}
\bibinfo{author}{\bibfnamefont{K.}~\bibnamefont{Svensson}},
  \bibinfo{author}{\bibfnamefont{Y.}~\bibnamefont{Jompol}},
  \bibinfo{author}{\bibfnamefont{H.}~\bibnamefont{Olin}}, \bibnamefont{and}
  \bibinfo{author}{\bibfnamefont{E.}~\bibnamefont{Olsson}},
  \bibinfo{journal}{Rev. Sci. Instr.} \textbf{\bibinfo{volume}{74}},
  \bibinfo{pages}{4945} (\bibinfo{year}{2003}).

\bibitem[{\citenamefont{Treacy et~al.}(1996)\citenamefont{Treacy, Ebbesen, and
  Gibson}}]{Treacy:1996}
\bibinfo{author}{\bibfnamefont{M.~M.~J.} \bibnamefont{Treacy}},
  \bibinfo{author}{\bibfnamefont{T.~W.} \bibnamefont{Ebbesen}},
  \bibnamefont{and} \bibinfo{author}{\bibfnamefont{J.~M.}
  \bibnamefont{Gibson}}, \bibinfo{journal}{Nature}
  \textbf{\bibinfo{volume}{381}}, \bibinfo{pages}{678} (\bibinfo{year}{1996}).

\bibitem[{\citenamefont{Lee et~al.}(2006)\citenamefont{Lee, Ophus, Fischer,
  Nelson-Fitzpatrick, Westra, Evoy, Radmilovic, Dahmen, and Mitlin}}]{Lee:2006}
\bibinfo{author}{\bibfnamefont{Z.}~\bibnamefont{Lee}},
  \bibinfo{author}{\bibfnamefont{C.}~\bibnamefont{Ophus}},
  \bibinfo{author}{\bibfnamefont{L.~M.} \bibnamefont{Fischer}},
  \bibinfo{author}{\bibfnamefont{N.}~\bibnamefont{Nelson-Fitzpatrick}},
  \bibinfo{author}{\bibfnamefont{K.~L.} \bibnamefont{Westra}},
  \bibinfo{author}{\bibfnamefont{S.}~\bibnamefont{Evoy}},
  \bibinfo{author}{\bibfnamefont{V.}~\bibnamefont{Radmilovic}},
  \bibinfo{author}{\bibfnamefont{U.}~\bibnamefont{Dahmen}}, \bibnamefont{and}
  \bibinfo{author}{\bibfnamefont{D.}~\bibnamefont{Mitlin}},
  \bibinfo{journal}{Nanotechn.} \textbf{\bibinfo{volume}{17}},
  \bibinfo{pages}{3063} (\bibinfo{year}{2006}).

\bibitem[{\citenamefont{Luber et~al.}(2008)\citenamefont{Luber, Mohammadi,
  Ophus, Lee, Nelson-Fitzpatrick, Westra, Evoy, Dahmen, Radmilovic, and
  Mitlin}}]{Luber:2008}
\bibinfo{author}{\bibfnamefont{E.}~\bibnamefont{Luber}},
  \bibinfo{author}{\bibfnamefont{R.}~\bibnamefont{Mohammadi}},
  \bibinfo{author}{\bibfnamefont{C.}~\bibnamefont{Ophus}},
  \bibinfo{author}{\bibfnamefont{Z.}~\bibnamefont{Lee}},
  \bibinfo{author}{\bibfnamefont{N.}~\bibnamefont{Nelson-Fitzpatrick}},
  \bibinfo{author}{\bibfnamefont{K.}~\bibnamefont{Westra}},
  \bibinfo{author}{\bibfnamefont{S.}~\bibnamefont{Evoy}},
  \bibinfo{author}{\bibfnamefont{U.}~\bibnamefont{Dahmen}},
  \bibinfo{author}{\bibfnamefont{V.}~\bibnamefont{Radmilovic}},
  \bibnamefont{and} \bibinfo{author}{\bibfnamefont{D.}~\bibnamefont{Mitlin}},
  \bibinfo{journal}{Nanotechn.} \textbf{\bibinfo{volume}{19}},
  \bibinfo{pages}{125705} (\bibinfo{year}{2008}).

\bibitem[{\citenamefont{Nilsson et~al.}(2004)\citenamefont{Nilsson,
  Borris{\'e}, and Montelius}}]{Nilsson:2004}
\bibinfo{author}{\bibfnamefont{S.~G.} \bibnamefont{Nilsson}},
  \bibinfo{author}{\bibfnamefont{X.}~\bibnamefont{Borris{\'e}}},
  \bibnamefont{and}
  \bibinfo{author}{\bibfnamefont{L.}~\bibnamefont{Montelius}},
  \bibinfo{journal}{Appl. Phys. Lett.} \textbf{\bibinfo{volume}{85}},
  \bibinfo{pages}{3555} (\bibinfo{year}{2004}).

\end{thebibliography}

\end{document}